# SEX DETECTION IN THE EARLY STAGE OF FERTILIZED CHICKEN EGGSVIA IMAGE RECOGNITION

Ufuk Asil and Efendi Nasibov

Department of Computer Science, Dokuz Eylul University, Izmir, Turkey

## ABSTRACT

*Culling newly hatched male chicks in industrial hatcheries poses a serious ethical problem. Both laying and broiler breeders need males, but it is a problem because they are produced more than needed. Being able to determine the sex of chicks in the egg at the beginning or early stage of incubation can eliminate ethical problems as well as many additional costs. When we look at the literature, the methods used are very costly, low in applicability, invasive, inadequate in accuracy, or too late to eliminate ethical problems. Considering the embryo's development, the earliest observed candidate feature for sex determination is blood vessels. Detection from blood vessels can eliminate ethical issues, and these vessels can be seen when light is shined into the egg until the first seven days. In this study, sex determination was made by morphological analysis from embryonic vascular images obtained in the first week when the light was shined into the egg using a standard camera without any invasive procedure to the egg.*

## 1. INTRODUCTION

If the chicks hatched in industrial hatcheries are male, the surplus of male chicks results in culling. It is a serious ethical problem to annihilate male chicks in hatcheries when they are only one day old. In Germany alone, approximately 42 million male chicks are culled annually. In Europe, 300 million overall. In France, with the decree published in the Official Gazette in 2022, killing the male chicks hatched from the eggs in the hatcheries that produce laying hens is prohibited. The law states that eggs can be killed by selection in the first 15 days of the 21-day incubation period, and any culling after 15 days is prohibited [1]. In Germany, a similar law will be enacted in 2024, banning embryo culling after the first seven days [2]. It is foreseen that such laws will become widespread daily, especially in Europe. In commercial hatcheries, eggs can be used as food when sexing is known before the eggs are incubated. This does not seem possible without an invasive method. Determining the sex in the early hatching stage can partially eliminate the ethical problems and additional costs. Eggs of male embryos can be used as an alternative feed raw material or as ingredients for industrial products.

## 2. LITERATURE REVIEW

When we look at the literature, the methods used for sex determination in chicks can be divided into two: classical and modern [3]. Classical methods include looking at the cloaca of the chicks after hatching and looking at the color and shape of the feathers. Applying classical methods can create severe economic expenses, such as occupying space in the incubators, energy costs, and trained personnel, considering that half of the eggs in the facilities are male. Modern methods are accepted, such as measuring estrogen from allantoic fluid, observing vessels, detection of ZZ/ZW chromosomes, and feather color analysis of the embryo. These methods could not be commercialized industrially due to the high cost, partial damage to the egg, low accuracy, or late-stage sex determination. In addition, invasive methods such as needle insertion and perforation cause low hatching rates [4].

## 2.1. Pre-Incubation Methods

The Magnetic Resonance Imaging (MRI) technique can determine the exact location of the blastodermin the yolk [5]. Sex gene sequences can be obtained from 4-400 cells in the blastoderm by biopsy, and sex can be determined by PCR[6][7] With flow cytometry, cells in a suspension are passed through a chamber illuminated by laser light. The analysis is done by evaluating the fluorescent signals cells give as they pass in front of the light. Feedback may be the physical properties of the cell, such as size, as well as the information on the various fluorochromes that bind to the cell. Thus, information about multiple properties, such as cell DNA content, enzyme activities, cell membrane potential, and viability, can be collected [8] [9].

In some studies, it has been determined that there is a low correlation between the shape index of Branta canadensis goose eggs and embryo sex [10].

## 2.2. Incubation Methods

The level of estrogen hormone from the mother decreases in the first developmental stage of the embryo in the egg, and each sex produces its hormone as the embryo develops. Hormones differ in growing embryos [11]. Sex can be determined by taking 20 μl of allantoic fluid from the incubated eggs (between 15-17 days) and analyzing oestradiol by the radio-immune method. Studies are ongoing to make this method commercially available [12] [13].

In another method developed by Weissmann et al., oestradiol, estrone sulfate, and testosterone levels in the allantoic fluid are analyzed by enzyme immunoassay Elisa method on the 9th day of incubation.This method can be used for gender determination at a rate of 98% [14]. However, this method is complex and invasive since a needle is inserted into the egg. Since the test results can be obtained in about 4 hours, its applicability in industrial poultry is low.

Ultrasonography gets information by reflecting high-frequency sound waves (ultrasound) applied to incubated eggs from different tissue surfaces. Using real-time B-mode ultrasonography, the embryo's morphological structure, and developmental stage in the egg can be monitored every period [15].

The number of heartbeats may differ according to gender in the embryo [16]. It isn't easy to monitor the heartbeat in embryos before the 15th day. Studies have shown that, as in adults, the heart rate of females in embryos between 15 and 20 days is different from that of males. Although the study is old, it is inspiring as the beginning of modern methods.

Researchers at the University of Dresden, Germany, reported that using spectroscopic techniques, gender can be determined with an accuracy of 95% by studies based on the embryo's blood vessels in the egg [4]. On the 3rd day of embryo development, blood vessels begin to form, while nerve cells are not functional [17]. For this reason, the researchers idealized the 3rd day ethically in their studies.In the method, a small hole was drilled in the upper part of the egg without damaging the membrane, blood vessels and heart on the yolk were monitored, and the biochemical properties of the embryo vessels were visualized with a spectrometer. Researchers claim that 3.5-day-old male and female embryos can be classified with the help of computers [4].

In this method, the researchers stated that they could distinguish between sexes with 90% success by performing spectrum analysis in the blood circulating in the vessels of the outer part of the embryo [4]. The first day that the feathers of the embryos appear during the incubation period is the 9th day. Brown chickens are one of the dominant breeds in industrial chicken breeding farms. Gender discrimination can be made by looking at the color of thefeathers of these chickens after hatching. Studies claim that 95% of sex determination is made between 11-14 days of incubation with hyperspectral imaging [18]. It has not been foreseen that this gender determination method could be performed in all genders.

## 2.3. Post-incubation methods

The vent method, one of the industry's most widely used traditional methods, was first described by Japanese scientists [19]. This method, in which 1-day-old chicks are selected by looking at the cloaca, is also known as the Japanese technique. Experts who receive serious training are expected to discriminate between 800-1200 chicks per hour with 1-3% error to meet industrial standards.

Gender discrimination can be made by using the length of the feathers on the wings of a few days-old chicks [20] [21]. This feathering rate in chicks is determined by genes related to sex. Slow pubescence dominates fast pubescence [22]. Therefore, the difference in length between the wing feathers and cover feathers of a few day-old chicks can be easily distinguished. By using this feature, gender discrimination in many races can be made widely [23].

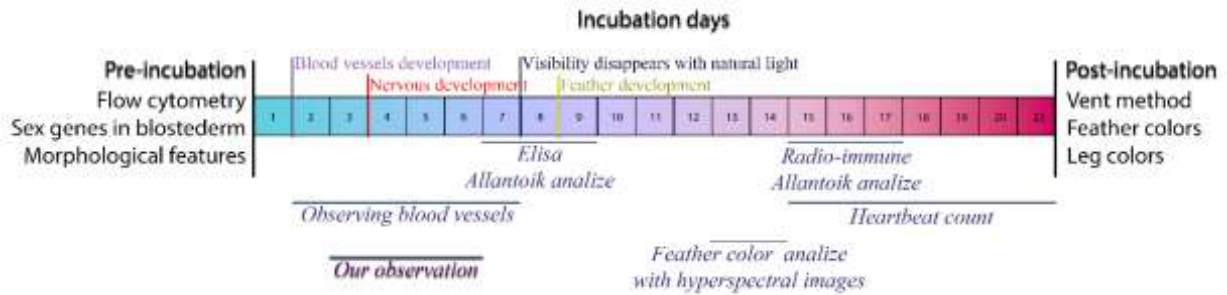

Figure 1. Chick sex determination methods in the literature

In Figure 1, the primary methods in the literature are visualized. Considering the methods, our method is an excellent candidate to eliminate ethical problems as it does not harm the egg and can only be observed in the first stage of the egg. Currently, various devices called ovoscopes are used in commercial hatcheries to check for anomalies in the egg and whether the egg is fertilized. The method we have developed provides an additional gain rather than an additional expense, as it can make all the diagnostics made by the ovoscope based on artificial intelligence.

## 3. MATERIAL AND METHODS

This study used broiler and Atak-S chicken eggs, the most used chicken breeds in industrial egg production. As seen in Figure 2a, the eggs were examined in 2 different incubation periods, with approximately 120 eggs in each incubation. Two 200-lumen LED light sources were given from two opposite poles to the inside of the eggs. LED light sources that emit less heat have been chosen because high heat can potentially damage embryo development in the egg. The light was turned on only during data collection so that the egg did not encounter high temperatures. Photos were taken on days 3,5 and 7 by rotating it in 10 steps to make a full tour.

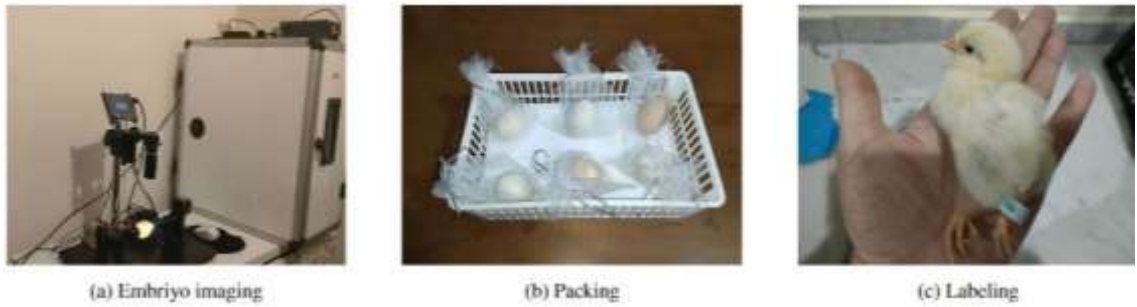

Figure 2. Data collection process

As the eggs are numbered on the 18th day of hatching, as seen in Figure 2b, the eggs were placed in pouches so the chicks would not mix. The chicks hatched, as seen in Figure 2c, are renumbered with bracelets.

## 4. IMAGE PROCESSING

It is well-known that image processing techniques in computer vision improve quality in applications such as classification. When we look at the literature, as seen in Figure 3, the embryo image in the first seven days of incubation is very similar to the eye image used for the diagnosis of Diabetic Retinopathy[24]. Studies on sex determination from hatching eggs are still in their infancy, and there is no pioneering method in the literature. A field of study investigating how chick embryos' vascular structures in broods differ according to sex has not yet been developed. For this reason, in this study, some morphological processes were applied sequentially, as in Diabetic Retinopathy, to increase the visibility of embryo vessels as features.

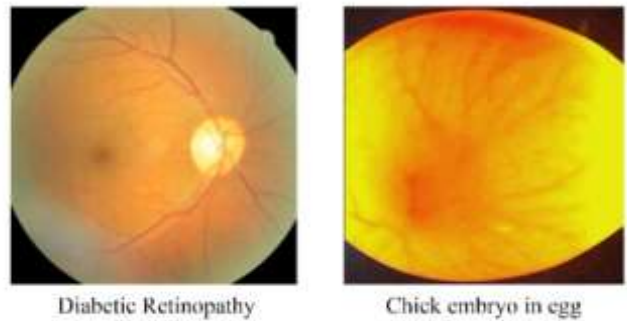

Figure 1. Similarity between Diabetic Retinopathy diagnosis and Incubation Embryo image[24]

As shown in Figure 4 below, it aims to reduce unnecessary data input into the system in the photos obtained from the egg. Since egg sizes vary a lot, a threshold was applied on the grey image, and cropping was applied to all photos with this data. When we look at the overall photos of the embryo in the egg, one of the most critical problems is the inability to visualize the vessels originating from the eggshell. Although this blur effect seems to be a problem initially, the Gaussian blur effect can have a positive impact when used to extract data from the picture using methods such as Adaptive GaussianThresholding [25] [26]. Unfortunately, there is no study investigating the effect of the blur effect of eggshells on sex determination.

International Journal of Computer Science & Information Technology (IJCSIT) Vol 15, No 2, April 2023

Histogram equalization (HE); aims to increase the features by spreading the region where the values in the picture are most intense to its neighbors, that is, by extending the density range. Contrast Limited Adaptive histogram equalization (CLAHE) unlike histogram equalization (HE), performs an equalization by dividing the image into blocks instead of the overall photos. Looking at the incubated eggs, it is seen that there are many different variable contrasts in vascularization. J.A. Stark clearly stated that contrast differences in vessel imaging with CLAHE eliminate the disadvantages of HE [27].

As shown in Figure 4, increasing the contrast with the Contrast Limited Adaptive histogram equalization (CLAHE) method is helpful for better visualization of vessels [27].

It is known that some noise may occur in the image with Histogram equalization, as shown in Figure 4. To reduce this noise effect, it is beneficial to apply Median blur to the photo [28].

The blood vessels were segmented, as shown in Figure 3 using the threshold So that The blood vessels are slightly more visible than the rest of the image. It is obvious that highlighting the features of segments carrying information in biological classifications will increase the success rate. By following the flowchart shown in Figure 4, the picture in which noise was removed with the median blur effect and the segmented picture of the veins were combined to maximize the meaningful data in the picture [29].

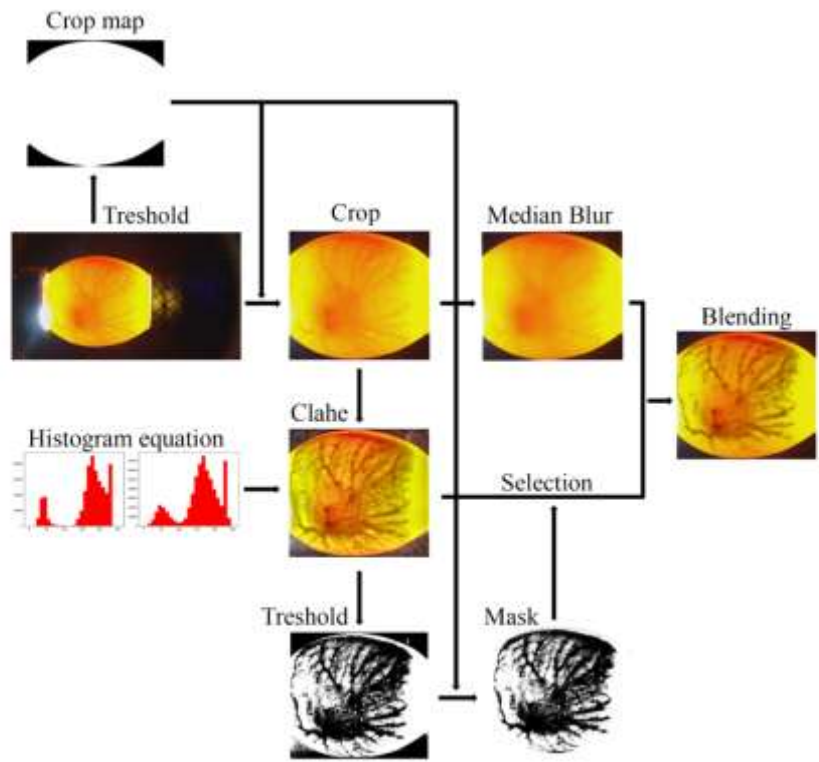

Figure 4. Image processing techniques

## 5. DATA ANALYSIS

While small scores were obtained that would not make sense in classifications made without applying any image processing method, successful results were obtained in gender determination in the incubation early stage after image processing. 112 photos independent of training photos were taken from Atak-S and Broiler eggs on the 3,5 and 7 days of incubation and were classified using the Resnet-50 [30], Vgg-16 [31], and Inception-V4 [32] models. The results were obtained as in Table 1.

Table 1
Classification results of different models of images of fertilized chicken eggs obtained on days 3,5 and 7.

| Incubation Time | Resnet-50 | Vgg-16 | Inception-V4 |
|---|---|---|---|
| 3 days | 62,75% | 65,69% | 67,65% |
| 5 days | 67,65% | 70,59% | 75,49% |
| 7 days | 69,61% | 72,55% | 72,55% |

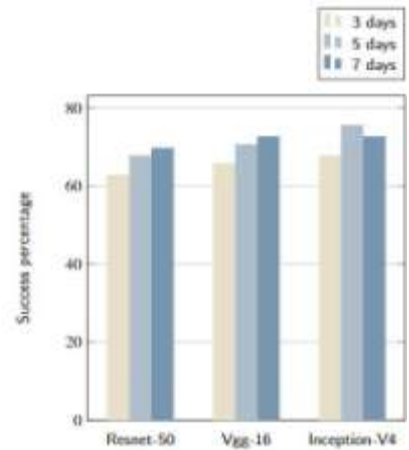

After 7 days, the images obtained from the fertilized egg in the incubator were not considered as early stage and because the light transmittance of the egg was visibly reduced, no classification was made. As can be seen from Table 1, the clarification of veining with image processing in high-resolution photographs of fertilized eggs in the early hatching stage makes it partially possible to determine the sex of the hatched chicks.

As seen in Table 1, the highest score belongs to the classification algorithm of the Inception V4 model on the 5th day. This result shows that the light-operated cameras, which are related to the loss of light permeability of the fluid in the egg, can reveal the sex feature on the 5th day at best.

## 6. CONCLUSION

Considering the laws enacted by European countries regarding the culling of chicks, methods of sex determination in chicks will change radically around the world. In parallel with this, when we look at the literature, it is clear that pre-incubation and early-stage sex determinations are permanent solutions. The main reason why most of these methods are not commercialized is the cost, technical difficulty, and time factor. Our study is very open to improvement as it is the first study in the literature. Although our study did not have a high success rate, providing vascularization of the embryo close to the eggshell by methods such as obtaining the image from the bottom or not moving the egg for the first three days may allow us to get a clear and stable vascular picture in each egg. Likewise, since the data is not received with expensive industrial equipment, the method can be functional even with a mobile phone outside the industry. More stable methods can be developed by combining them with various methods.

International Journal of Computer Science & Information Technology (IJCSIT) Vol 15, No 2, April 2023

International Journal of Computer Science & Information Technology (IJCSIT) Vol 15, No 2, April 2023

## AUTHORS

**Ufuk Asil** is a Ph.D. student at the Department of Computer Science, Dokuz Eylul University, Izmir, Turkey. He has been a Police officer since 2010. He received his master's degree in Nanoscience and Nanoengineering from Dokuz Eylul University. His research interests include Image Processing and Neuromorphic chip design. My personal website

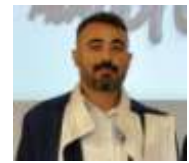

**Efendi Nasibov (Nasiboglu )** received a B.Sc. and M.Sc. Degrees in the Applied Mathematics Department from Baku State University, Azerbaijan, and a Ph.D. degree in Mathematical Cybernetics (Moscow) and a Dr.Sc. of Computer Science degree from the Institute of Cybernetics of the Academy of Science of Azerbaijan. He is currently a full Professor at the Department of Computer Science, Dokuz Eylul University, Izmir, Turkey. His research interests are in the application of Fuzzy Modelling, Data Mining, and AI techniques in Decision-Making problems.

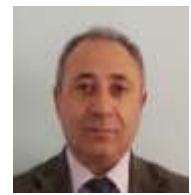